\documentclass[%
 reprint,
 showpacs,preprintnumbers,
 amsmath,amssymb,
 aps,
]{revtex4-1}

\usepackage{graphicx}
\usepackage{dcolumn}
\usepackage{bm}
\usepackage{braket}
\usepackage[version=3]{mhchem}

\newcommand{\figref}[1]{Fig.~\ref{#1}}

\newcommand{\mrm}{\mathrm}

\newcommand{\mcl}{\mathcal}

\newcommand{\lag}{\langle}
\newcommand{\rag}{\rangle}
\newcommand{\br}[1]{\left( #1 \right)}

\newcommand{\BR}[1]{\left[ #1 \right]}
\newcommand{\ev}[1]{\left\langle #1 \right\rangle}
\newcommand{\av}[1]{\left| #1 \right|}

\newcommand{\hspc}{\hspace{1em}}

\newcommand{\alp}{\alpha}
\newcommand{\bet}{\beta}

\newcommand{\del}{\delta}

\newcommand{\tht}{\theta}

\newcommand{\Del}{\Delta}

\newcommand{\ham}{\mcl{H}}

\newcommand{\vS}{\bm{S}}
\newcommand{\rnc}{\mrm{nc}}
\newcommand{\rest}{\mrm{est}}
\usepackage{color,ulem}

\begin{document}

\title{Classical and Quantum Magnetic Ground States on an Icosahedral Cluster}

\author{Shintaro Suzuki}
\affiliation{Department of Material Science and Technology, Tokyo University of Science, Katsushika, Tokyo 125-8585, Japan}
\author{Ryuji Tamura}
\affiliation{Department of Material Science and Technology, Tokyo University of Science, Katsushika, Tokyo 125-8585, Japan}
\author{Takanori Sugimoto}
\email{sugimoto.takanori@rs.tus.ac.jp}
\affiliation{Department of Applied Physics, Tokyo University of Science, Katsushika, Tokyo 125-8585, Japan}

\date{\today}

\begin{abstract}
Recent discovery of various magnetism in Tsai-type quasicrystal approximants, in whose alloys rare-earth ions located on icosahedral apices are coupled with each other via the Ruderman--Kittel--Kasuya--Yosida interaction, opens an avenue to find novel magnetism originating from the icosahedral symmetry. Here we investigate classical and quantum magnetic states on an icosahedral cluster within the Heisenberg interactions of all bonds. Simulated annealing and numerical diagonalization are performed to obtain the classical and quantum ground states. We obtain qualitative correspondence of classical and quantum phase diagrams. Our study gives a good starting point to understand the various magnetism in not only quasicrystal approximants but also quasicrystals.
\end{abstract}

\pacs{Valid PACS appear here}
\maketitle

\section{Introduction}
Long-range magnetic orders in quasi-periodic lattice have been fascinating and challenging targets since the discovery of quasicrystals~\cite{Shechtman84}. 
First investigation of magnetism in quasicrystals was performed in Al-Mn based alloy in 1986~\cite{Fukamichi86,Warren86,Hauser86}. 
Next, Bergman-type quasicrystals were examined thanks to the discovery of Zn-Mg-RE quasicrystals~\cite{Luo93,Tsai94} (RE = rare earth).
However, no magnetic long-range ordering has been observed in these qusicrystals so far~\cite{Hattori95,Sato98}. 
On the other hand, another quasicrystal with containing rare-earth elements was discovered by A. P. Tsai in 2000~\cite{Tsai00,Guo00,Guo00-2,Guo01}. 
To attain the targets, at present, Tsai-type quasicrystals are investigated~\cite{Sato01}, due to a key observation of antiferromagnetism in an approximant \ce{Cd6Tb}~\cite{Tamura10,Kim12,Mori12}, which have the same local structure as the quasicrystals but have a periodicity.

The Tsai-type quasicrystal approximants in common with the quasicrystals, consist of rhombic triacontahedral clusters. 
The cluster includes a concentric tetrahedron, a dodecahedron, an icosahedron, and an icosidodecahedron from center out, whose apices constituent ions are located on~\cite{Gomez03}. 
Among polyhedra, the rare-earth ions placed on the icosahedron only contributes magnetism helped by so-called the Ruderman--Kittel--Kasuya--Yosida (RKKY) interaction~\cite{Ruderman54,Kasuya56,Yosida57}, resulting in novel magnetic orders, e.g., multifarious magnetism discovered in Tsai-type 1/1 approximants~\cite{Ishikawa16,Ishikawa18,Miyazaki20}. 
Interestingly, the magnetism is controlled by constitutional ratio of ions in ternary alloys of the approximants via electron density, because the RKKY interaction depends on the Fermi wavenumber which is a function of the electron density in Fermi gas approximation. 
Actually, the Curie--Weiss temperature observed in the 1/1 approximants shows an oscillation as a function of estimated electron density indicating the RKKY interaction~\cite{Ishikawa16,Ishikawa18}.

Surprisingly, the latest experimental study on a Tsai-type 2/1 approximants has reported almost the same behaviors of magnetism as the 1/1 approximants despite difference of crystal structure between 1/1 and 2/1 approximants~\cite{Yoshida19,Inagaki20}. 
This result suggests importance of the common local structure, i.e., the rhombic triacontahedral cluster including the magnetic icosahedron. 
The numerical calculations of the magnetic ground state and physical properties in a single icosahedral cluster both in the classical and quantum Heisenberg model are already performed with nearest-neighbor exchange interaction~\cite{Konstantinidis05,Axenovich01,Hucht11}. 
However, these studies do not consider the interaction between the 2nd and 3rd neighbor spins. 
Therefore, we investigate magnetism of an isolated icosahedron within the 2nd and 3rd neighbor interactions. 
Especially, we focus our examination on the magnetic ground states to understand low-temperature physics in the Tsai-type approximants. 
Since magnitude of magnetic moment depends on rare-earth ions, we consider both quantum and classical spins corresponding to small and large magnetic moments, respectively.

\section{Model and Method}
\begin{figure}[tbp]
 \begin{center}
  \includegraphics[width=0.48\textwidth]{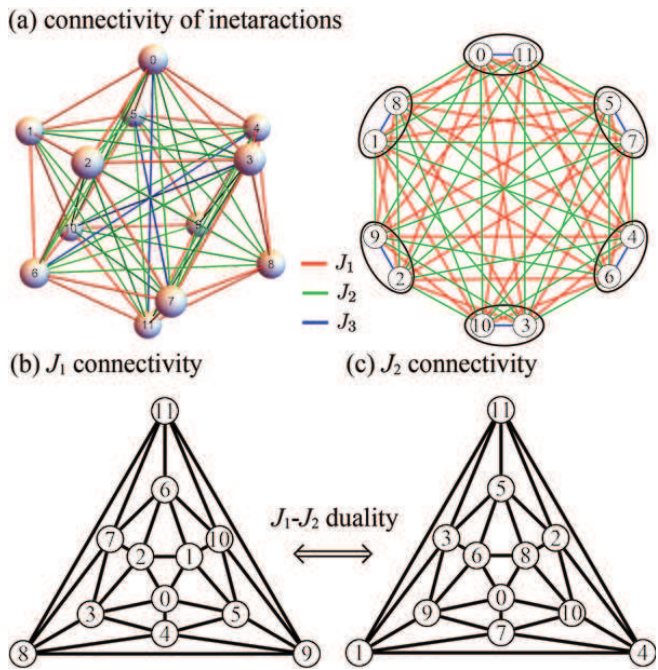}
 \end{center}
 \caption{(a) Schematic icosahedral spin cluster and its connectivity of interactions. 
Balls (circles) denote apices assigning spins. 
The red, green, and blue bonds correspond to 1st, 2nd, and 3rd neighbor interactions. 
An ellipse represents a pair of spins located on opposite apices of icosahedron. 
(b), (c) Connectivity of 1st or 2nd neighbor interaction.   
These graphs are equivalent.
}
 \label{fig1}
\end{figure}

In this paper, we examine magnetic ground states in the following model Hamiltonian,
\begin{equation}
\ham=\sum_{n=1}^3J_n\sum_{\lag i,j\rag_n} \vS_i\cdot\vS_j
\end{equation}
where $J_n$ and $\lag i,j\rag_n$ ($n=1,2,3$) represent the exchange energy and $n$-th neighbor bonds, respectively. 
For simplicity, instead of $J_n$, we use other angle parameters $\tht_J$ and $\phi_J$ to control the exchange energies $J_n$ via $(J_1,J_2,J_3)=J(\sin\tht_J\cos\phi_J,\sin\tht_J\sin\phi_J,\cos\tht_J)$ with the energy unit $J=\sqrt{J_1^2+J_2^2+J_3^2}=1$. 
The spin degree of freedom $\vS_i$ on $i$-th site is regarded as a unit vector in the classical model and a spin-$1/2$ operator in the quantum model. 
The spin sites are located at apices of an icosahedron [see \figref{fig1}(a)]. Concretely, we consider 12 spins with 30 bonds of 1st neighbor interaction, 30 bonds of 2nd neighbor interaction, and 6 bonds of 3rd neighbor interaction. Except for the 3rd neighbor interactions, connectivity of the interactions composes a complete hexapartite graph $K_{2,2,2,2,2,2}$. 
Figure 1(b)-(c) shows connectivity of the 1st and 2nd neighbor interactions. 
Since these graphs are the same, there is symmetry with respect to permutation of the site indexes corresponding to \figref{fig1}(b)-(c); i.e., background physics of two different points in the parameter space $(\tht_J, \phi_J)=(\tht_J, \pi/4-\del)$ and $(\tht_J, \pi/4+\del)$ are essentially equivalent, while the spin configurations are different at a glance. Such the correspondence between two different models is often called duality. Hence, we call the correspondence $J_1$-$J_2$ duality.

To obtain the ground state, we numerically apply simulated annealing method to the classical model and exact diagonalization method to the quantum model. 
The simulated annealing is a Monte-Carlo method where vector spins are updated one by one with a certain probability based on the statistical mechanics at each temperature. 
The temperature is gradually lowered to zero like the annealing process in heat treatment of real materials. 
On the other hand, the exact diagonalization is genuinely a quantum method at zero temperature to take quantum fluctuations into account. 
In this method, a matrix form of the Hamiltonian represented in the basis of spin wavefunctions is exactly diagonalized to obtain the eigenstate of the minimal energy.
We have confirmed accordance of ground-state energies obtained by the numerical methods and analytical energies at exactly solvable points in the parameter space, e.g., $(\tht_J, \phi_J)=(0, 0), (\pi, 0), (\tan^{-1}\sqrt{2}, \pi/4)$, and so on.

\section{Classical Magnetic States}
\subsection{Simmulated annealing}
In this section, we first show numerical results in the classical model with $\av{\vS_i}=1$. In this case, the Hamiltonian can be described as
\begin{equation}
  \ham=\sum_{n=1}^3J_nN_n\ev{\cos\alp_{ij}}_{\lag i,j\rag_n} 
\end{equation}
where $N_n$ and $\alp_{ij}$ denote the number of $n$-th neighbor bonds and an angle of two spin vectors $\vS_i$ and $\vS_j$, respectively. 
The bracket $\ev{\cos\alp_{ij}}_{\lag i,j\rag_n}$ represents mean value of inner product $\vS_i\cdot\vS_j$ connected with the $n$-th neighbor bonds, $N_n^{-1}\sum_{\lag i,j\rag_n}\cos\alp_{ij}$.

\begin{figure}[tbp]
 \begin{center}
  \includegraphics[width=0.48\textwidth]{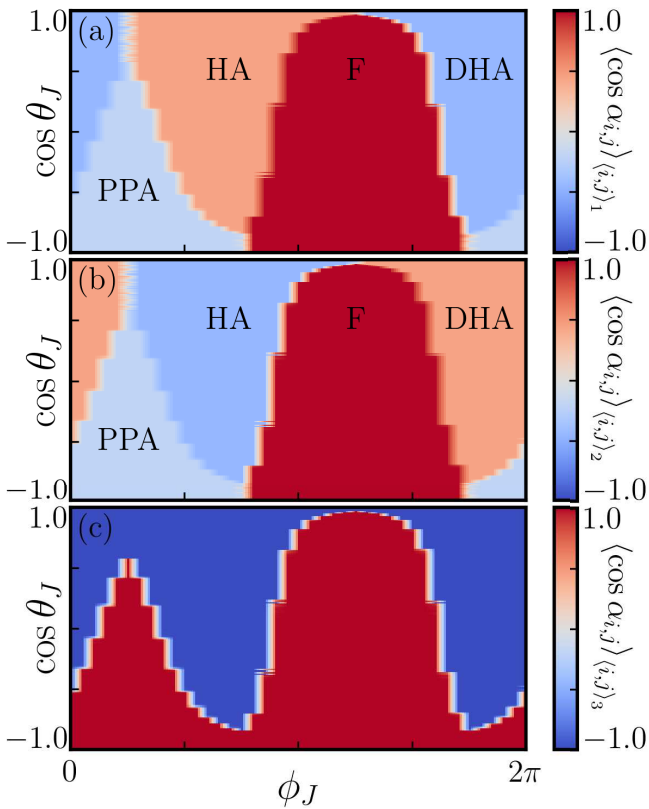}
 \end{center}
 \caption{Numerical results of the classical model: (a) $\ev{\cos\alp_{ij}}_{\lag i,j\rag_1}$, (b) $\ev{\cos\alp_{ij}}_{\lag i,j\rag_2}$ and (c) $\ev{\cos\alp_{ij}}_{\lag i,j\rag_3}$. 
In the parameter space, we take $41\times 201$ sample points for calculation. See \figref{fig3} for the acronyms, HA, DHA, PPA, and F.
 }
 \label{fig2}
\end{figure}

Figure 2(a)-(c) shows the $\ev{\cos\alp_{ij}}_{\lag i,j\rag_n}$ in the parameter space $(\tht_J, \phi_J)$ obtained by the simulated annealing. 
The mean values $\ev{\cos\alp_{ij}}_{\lag i,j\rag_1}$ and $\ev{\cos\alp_{ij}}_{\lag i,j\rag_2}$ take only four values $-0.44, -0.20, 0.44$ and $1.00$ in \figref{fig2}(a)-(b). 
In consideration of these mean values, the spin configuration can be classified principally into four ground-state phases. 
Besides, $\ev{\cos\alp_{ij}}_{\lag i,j\rag_3}=\pm 1$ in \figref{fig2}(c) indicates only two configurations of spins on the opposite sites of icosahedral apices, corresponding to parallel and anti-parallel spins.

\begin{figure*}[tbp]
 \begin{center}
  \includegraphics[width=0.7\textwidth]{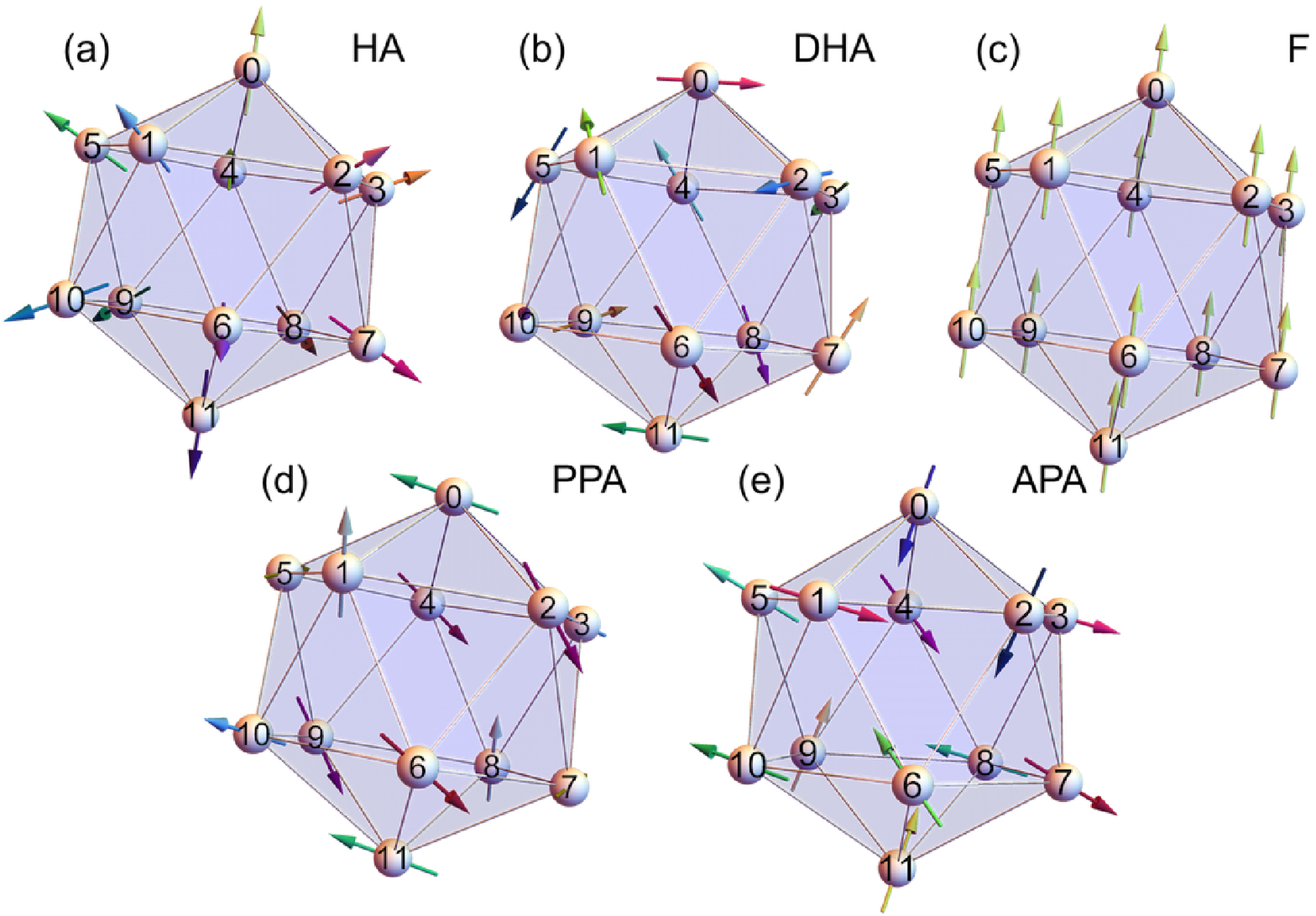}
 \end{center}
 \caption{(a)-(d) Examples of ground-state spin configuration in hedgehog antiferromagnetic (HA), dual hedgehog antiferromagnetic (DHA), ferromagnetic (F), parallel pairs' antiferromagnetic (PPA) phase. 
(e) Spin configuration of antiparallel pairs' antiferromagnetic (APA) state appearing at the boundary of the HA and DHA phases.
 }
 \label{fig3}
\end{figure*}

Figure 3(a)-(e) displays the five spin configurations in an icosahedral cluster obtained by the numerical calculation. 
Characteristics of these configurations are listed below. Note that the energy is invariant with respect to any global O(3) rotation.
\begin{enumerate}
\item[(a)] Hedgehog antiferromagnetic (HA) state: spin vectors are parallel to the normal vectors of circumscribed sphere of icosahedron like a hedgehog. 
\item[(b)] Dual hedgehog antiferromagnetic (DHA) state: a conjugate state of the HA state with respect to site exchange of the $J_1$-$J_2$ duality in \figref{fig1}(b) and (c).
\item[(c)] Ferromagnetic (F) state: all spins are parallel.
\item[(d)] Parallel pairs' antiferromagnetic (PPA) state: spins located on opposite apices of icosahedron are parallel, whereas total moment is zero.
\item[(e)] Antiparallel pairs' antiferromagnetic (APA) state: spins located on opposite apices of icosahedron are antiparallel.
\end{enumerate}

In the following, we explain these states in detail.
In \figref{fig3}(a), since two spins with a 3rd neighbor bond are antiparallel, total spins are cancelled, indicating antiferromagnetism. 
In addition, mean values of the 1st and 2nd neighbor inner products imply a distinct order in \figref{fig2}(a)-(b). 
More certainly, we find that the spin vectors in this phase correspond to the normal vectors of circumscribed sphere of icosahedron with applying an appropriate global O(3) rotation, which conserves all angles of spins and total energy. We thus call this hedgehog antiferromagnetic (HA) phase. 
Based on the hedgehog structure, we can estimate the inner product as follows
\begin{equation}
  \ev{\cos\alp_{ij}}_{\lag i,j\rag_1} = -\ev{\cos\alp_{ij}}_{\lag i,j\rag_2} = \frac{\tau}{\tau+2} = 0.447.., 
\end{equation}
where $\tau$ is the golden ratio. The spin configuration in the HA order is also discussed in Sec. 3.2. 
As mentioned in Sec. 2, there is a duality of the 1st and 2nd neighbor interactions with respect to permutation of the site indexes corresponding to \figref{fig1}(b)-(c). 
Reflecting this duality, the HA phase corresponds to the dual HA phase, and thus, we call this ordering as dual hedgehog antiferromagnetic (DHA) phase. 
Note that the HA phase is quite similar to the cuboc order reported in numerical study on the RKKY magnetism in Tsai-type 1/1 approximant~\cite{Miyazaki20}, whereas symmetry group of an icosahedron is basically different from that of cubic. 
Furthermore, we mention that although the DHA phase is also similar to the magnetic structure determined by neutron scattering on Au-Al-Tb 1/1 approximant~\cite{Sato19}, those should be in general discussed individually because of difference of the models.

 In the ferromagnetic (F) phase, $\ev{\cos\alp_{ij}}_{\lag i,j\rag_n}=1$ for all $n$ implies all spins are parallel as shown in \figref{fig3}(c). 
This numerical calculation does not reproduce the spin configuration of ferromagnetism determined by neutron diffraction in Au-Si-Tb 1/1 approximants~\cite{Hiroto20}. 
This can be an evidence that the anisotropy caused by total angular momentum of Tb ion dominates the low-temperature magnetism~\cite{Sugimoto16}, though we do not consider the anisotropy in this study.
In the parallel pairs' antiferromagnetic (PPA) phase, the spin configuration changes one by one in many trials of calculation, while two spins on opposite apices of icosahedron, which are connected with a 3rd neighbor bond, are always parallel [see \figref{fig3}(d)]. 
Therefore, we call this PPA phase. 
Undoubtedly, \figref{fig2}(a)-(b) displays that the same mean values $\ev{\cos\alp_{ij}}_{\lag i,j\rag_1} = \ev{\cos\alp_{ij}}_{\lag i,j\rag_2} =-0.20$, implying no distinct orders. 
In fact, if a certain spin’s angles to 1st and 2nd neighboring spins are completely random values and the mean values are the same, the mean value reads,

Similarly, at the boundary between the HA and DHA phases ($\phi_J=\pi/4, 5\pi/4$), the spin configuration varies one by one in trials of calculation. 
Contrary to the random values of $\ev{\cos\alp_{ij}}_{\lag i,j\rag_1} $ and $\ev{\cos\alp_{ij}}_{\lag i,j\rag_2}$, $\ev{\cos\alp_{ij}}_{\lag i,j\rag_3}=-1 $ is fixed at the boundary, indicating antiferromagnetism based on pairs of antiparallel spins on opposite apices of icosahedron. 
Hence, we call this antiparallel pairs' antiferromagnetic (APA) state. The appearance of this state should be caused by competition of the HA and DHA phases. 

\begin{figure}[tbp]
 \begin{center}
  \includegraphics[width=0.48\textwidth]{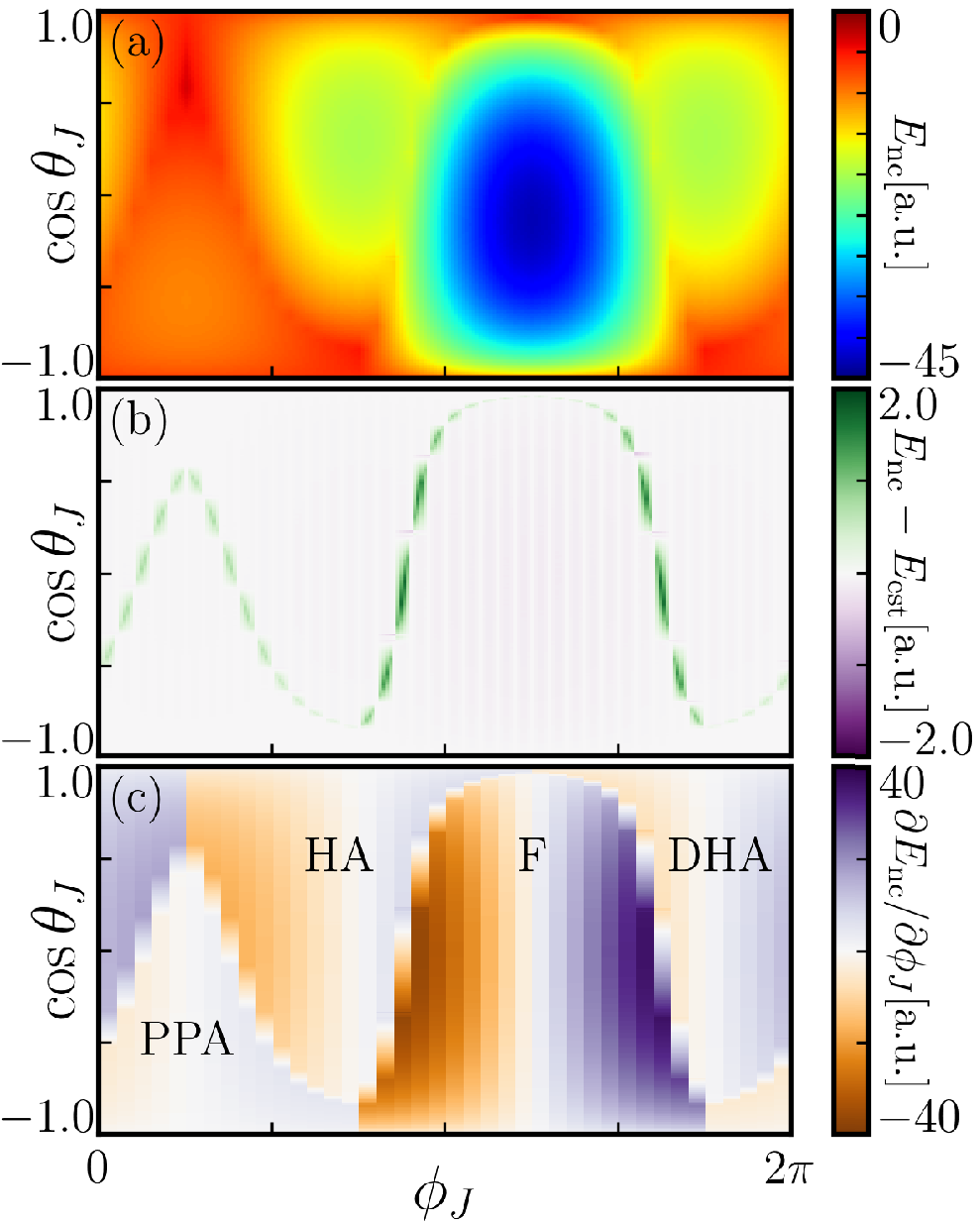}
 \end{center}
 \caption{(a) Numerical result of the total energy $E_{\rnc}$ in the classical model. (b) Energy difference between $E_{\rnc}$ and $E_{\rest}$ which is analytically estimated by the ground-state spin configurations in \figref{fig3}. (c) The derivative of $E_{\rnc}$ with respect to $\phi_J$.}
 \label{fig4}
\end{figure}

We show total energy $E_\rnc$ as a function of $(\tht_J, \phi_J)$ in \figref{fig4}(a), which is obtained numerically. 
To evaluate the $E_\rnc$, this result is compared with the energy function $E_\rest$, which is estimated by the spin configurations in four phases: $30[\tau/(\tau+2)](J_1-J_2)-6J_3$ for the HA phase, $30[\tau/(\tau+2)](J_2-J_1)-6J_3$ for the DHA phase, $30(J_1+J_2 )+6J_3$ for the F phase, and $-6(J_1+J_2 )+6J_3$ for the PPA phase. 
Note that the APA state appears only at $J_1=J_2$, so that this is merged into the HA and DHA phases. 
Energy difference $E_\rnc-E_\rest$ displayed in \figref{fig4}(b) gives a good coincidence between $E_\rnc$ and $E_\rest$ except for phase boundaries, where numerical accuracy is not enough. 
Furthermore, the phase boundaries are apparently obtained by the derivative of total energy with respect to $\phi_J$ in \figref{fig4}(c).

\subsection{Analytical explanations}
To explain the ground-state phase transitions in the classical model, we consider two specific conditions, (I) $J_1=J_2$ and (II) $J_1>0$ ($J_2>0$) with $J_2=J_3=0$ ($J_1=J_3=0$). 
Condition (I) corresponds to $\phi_J=\pi/4$ and $5\pi/4$ lines. Condition (II) is $(\tht_J, \phi_J)=(\pi/2, 0)$ or $(\pi/2, \pi/2)$.

{\it Condition} (I)--- At the symmetric line of $J_1=J_2$, the classical Hamiltonian corresponds to the following form,
\begin{equation}
  \ham=\frac{J_1}{2}\BR{\br{\sum_{i=0}^{11}\vS_i}^2-12}+(J_3-J_1)\sum_{\lag i,j\rag_3}\vS_i\cdot\vS_j.
\end{equation}
We can obtain the ground state in four cases, (i) $J_1>0$, $J_3>J_1$, (ii) $J_1>0$, $J_3<J_1$, (iii) $J_1<0$, $J_3<J_1$, and (iv) $J_1<0$, $J_3>J_1$. 
In case (i), which corresponds to $\tht_J<\tan^{-1}\sqrt{2}$ and $\phi_J=\pi/4$, the minimum energy is obtained with $\vS_i=-\vS_{\bar{i}}$, where $\bar{i}$ denotes the opposite site from $i$-th apex of icosahedron. 
Therefore, the ground state in case (i) is understood by antiferromagnetic based on six pairs of antiparallel spins connected by 3rd neighbor bond [see \figref{fig1}(a)], obtained as the APA state [also see \figref{fig3}(e) for the spin configuration]. 
This case however seems unstable and corresponds to a boundary between the HA and DHA phases obtained by numerical calculation [see \figref{fig3}(a), (b) and \figref{fig4}(a)]. 
The ground state in case (ii) is similar to case (i), whereas this case is stable in numerically-obtained phase diagram, corresponding to $\tht_J>\tan^{-1}\sqrt{2}$ and $\phi_J=\pi/4$ in \figref{fig2}(a)-(c) and \figref{fig3}(a). 
The minimum energy is obtained with $\vS_i=\vS_{\bar{i}}$ and $\sum_i\vS_i=0$, where spins on opposite apices of icosahedron are parallel under zero net moment of icosahedron, leading antiferromagnetism. 
Case (iii) is more trivial. Spin configuration of the minimum energy is ferromagnetism, i.e., $\vS_i=\vS_j$, corresponding to $\tht_J>\tan^{-1}\sqrt{2}$ and $\phi_J=5\pi/4$ in \figref{fig2}(a). 
In case (iv), assuming that angle of spins on opposite apices of icosahedron is $\alp$ and that every composite vector of the spins on opposite apices is the same, the energy is given by,
\begin{equation}
  E=\frac{J_1}{2}\BR{(12\cos\alp)^2-12}+6(J_3-J_1)\cos\alp
\end{equation}
With decreasing $J_3$, the ground state changes from antiferromagnetic ($\alp=\pi$) to ferromagnetic ($\alp=0$) at $J_3=-5J_1$ corresponding to $\tht_J>\tan^{-1}(\sqrt{2}/5)$ with $\phi_J=5\pi/4$. 
The antiferromagnetic state is also the APA, so that this antiferromagnetic state is also unstable and the boundary between the HA and DHA phases. 
On the other hand, the ferromagnetic state is merged into the phase of case (iii).

{\it Condition} (II)--- We first consider the ground state with only 1st neighbor interaction. 
In this condition, we assume following spin configuration with vector spins $\vS_i=(\sin\alp_i\cos\bet_i,\sin\alp_i\sin\bet_i,\cos\alp_i)$, $\alp_0=0$, $\bet_1=0$, and
\begin{equation}
  \alp_i =\alp_5\equiv \alp,\ \bet_{i+1}-\bet_{i} =\bet_1-\bet_5 \equiv \Del \ (\mrm{mod.}\>2\pi),
\end{equation}
for $i=1,2,\cdots,4$.
With the assumption, the exchange energy $E_{i,j}=\vS_i\cdot\vS_j$ $(i,j\leq 5)$ is given by
\begin{equation}
  E_{0,i}=\cos\alp,\ E_{i,i+1}=E_{5,1}=\cos^2\alp+\sin^2\alp\cos\Del
\end{equation}
 If these energies are the same, $\cos\alp=1, \cos\Del/(1-\cos\Del)$. In addition, since $\Del=2\pi n/5$ ($n=0,\pm 1, \pm 2$), the energy has one of three values 
 \begin{equation}
   E_{i,j}=\cos\alp =
   \begin{cases}
     1 & (\Del = 0) \\
     1/\sqrt{5} & (\Del=\pm 2\pi/5)\\
     -1/\sqrt{5} & (\Del=\pm 4\pi/5)
   \end{cases}.
 \end{equation}
Therefore, because of the antiferromagnetic interaction, the third value is chosen as the minimum energy. 
Interestingly, if we consider antiparallel spins on opposite apices of icosahedron, i.e., $\vS_i=-\vS_{\bar{i}}$, all exchange energies of 1st neighbor interaction are the same. 
This spin configuration gives a good accordance with the numerical result in the DHA phase [\figref{fig3}(b)]. Note that the spin configuration has global O(3) rotation degree of freedom. 
On the other hand, if there is only 2nd neighbor antiferromagnetic interaction, we can obtain the ground state by using the $J_1$-$J_2$ duality in \figref{fig1}(b)-(c). 
Starting from the spin configuration discussed above, the permutation of spin sites also gives a good coincidence with the numerical result in the HA phase [see \figref{fig3}(a)]. 
In fact, with an appropriate global O(3) rotation, the spin vector has the same direction as normal vector of circumscribed sphere of icosahedron, like a hedgehog. 

\section{Quantum Magnetic States}
In this section, we show numerical results in the quantum model with spin-$1/2$ operators located at twelve apices of icosahedron. 
The ground state only with antiferromagnetic 1st neighbor interaction has already been presented by N. P. Konstantinidis~\cite{Konstantinidis05}. 
However, effects of 2nd and 3rd neighbor interactions on the ground state remain unclear so far. 
Thus, we investigate these effects and compare quantum phase diagram with the classical one.

\begin{figure}[tbp]
 \begin{center}
  \includegraphics[width=0.45\textwidth]{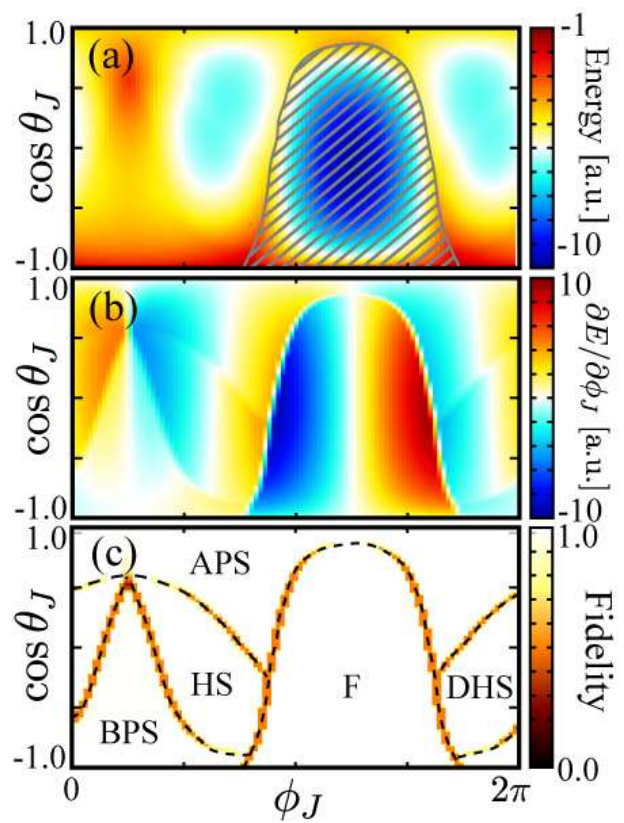}
 \end{center}
 \caption{(a) Ground-state energy and (b) its derivative with respect to $\phi_J$ obtained by exact diagonalization of the quantum model. 
In the parameter space, we take $100\times 100$ sample points for calculation. The shaded area in (a) represents $S_{\mrm{tot}}=6$, i.e., ferromagnetism. 
(c) Fidelity of the ground state. The dotted lines drawn by hand represents the phase boundaries among hedgehog singlet (HS), dual hedgehog singlet (DHS), bonding pairs' singlet (BPS), antibonding pairs' singlet (APS), and ferromagnetic (F) phases.}
 \label{fig5}
\end{figure}

We first calculate ground-state energies in the Hilbert subspace limited by $S_{\mrm{tot}}^z=\sum_i S_i^z$. 
By checking degeneracy of ground states between different subspaces, we can determine magnitude of total spin $S_{\mrm{tot}}$, e.g., if the ground-state energies of only $S_{\mrm{tot}}^z=0$ and $1$ are the same, the ground state is triplet ($S_{\mrm{tot}}=1$). 
Figure 5(a)-(b) shows the ground-state energy and its derivative with respect to $\phi_J$ in the parameter space. 
In the shaded area of \figref{fig5}(a), ground states are 13-fold degeneracy, i.e., $S_{\mrm{tot}}=6$ ground state corresponding to the ferromagnetic (F) state. 
Except for the shaded area, we have confirmed no degeneracy between different subspaces, so that singlet ground state appears. 
In the singlet area, we can see anomalous lines in \figref{fig5}(b), which imply phase boundaries. 
To confirm the phase boundaries, we also calculate an overlap of ground states with neighboring sample points in the parameter space, so-called fidelity, defined by
\begin{equation}
  \mrm{Fd}(\tht_J,\phi_J;\del)=\av{\lag \mrm{gs}:\tht_J,\phi_J|\mrm{gs}:\tht_J,\phi_J+\del \rag}.
\end{equation}
If the ground state is continuously deformed, that indicates no degeneracy at ground state and no phase transition between $(\tht_J, \phi_J)$ and $(\tht_J$,$\phi_J+\del)$, the fidelity converges to the unity $\lim_{\del\to 0}\mrm{Fd}(\tht_J,\phi_J;\del)\to 1$. 
Otherwise, the fidelity is much smaller than $1$. 
Therefore, we can determine the phase boundary by using the fidelity even in the singlet area. 
Figure 5(c) shows the fidelity, and we can see several lines with a dip of the fidelity, which corresponds to anomalous lines in \figref{fig5}(b). 
Thus, we conclude that there are four singlet phases except for the ferromagnetic phase.

The four singlet phases are understood as follows. 
The upper region of singlet phase includes the north pole $\tht_J=0$ and its ground state is intuitively described by that at the north pole. 
Since only the 3rd neighbor antiferromagnetic interaction is non-zero at the north pole, two spins on opposite apices of icosahedron compose a singlet (antibonding) pair and the ground state is the direct product of six singlet pairs. 
Hence, we call the upper region antibonding pairs' singlet (APS) phase. 
On the other hand, the south pole $\tht_J=\pi$ requires close attention because the south pole is a singular point between the ferromagnetic phase and the lower region of singlet phase. 
In fact, at the south pole, where $J_1=J_2=0$ and $J_3>0$, six triplet (bonding) pairs consisting of spins on opposite apices of icosahedron are completely decoupled. 
With slight positive $J_1=J_2>0$, which is included in the lower region of singlet phase, six triplets antiferromagnetically interact with each other, resulting in a singlet ground state. 
Thus, we call the lower region bonding pairs' singlet (BPS) phase. 
In middle region, there are two singlet phases more, which include parameter points with only the 1st and 2nd neighbor antiferromagnetic interactions, i.e., $(\tht_J, \phi_J)=(\pi, 0)$ and $(\pi/2, \pi/2)$, respectively. 
In the classical model, the ground state with only the 1st or 2nd neighbor interaction is discussed in Sec. 3.2. 
Especially, with only the 2nd neighbor interaction, the classical spin configuration is hedgehog-like. 
Therefore, we call these two phases hedgehog singlet (HS) and dual hedgehog singlet (DHS) phases, respectively.

\begin{figure}[tbp]
 \begin{center}
  \includegraphics[width=0.45\textwidth]{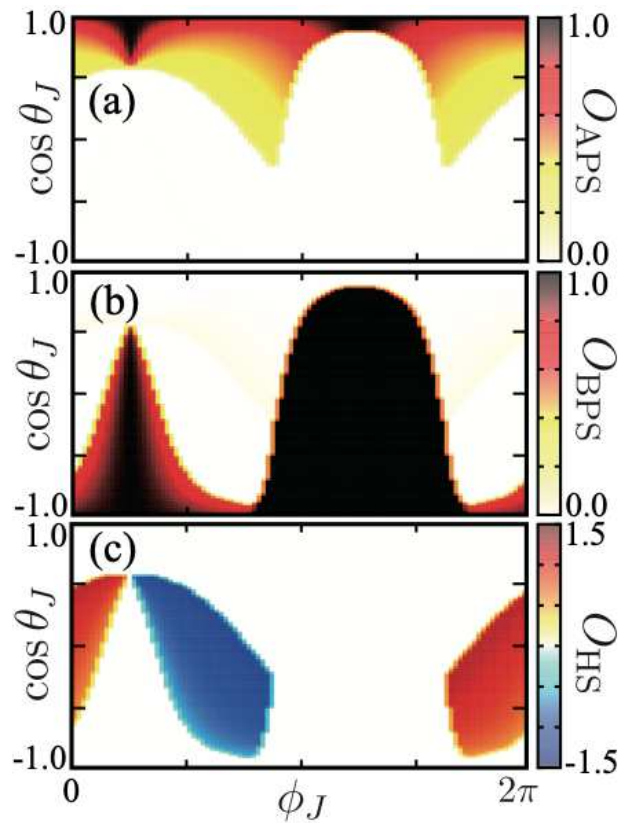}
 \end{center}
 \caption{Order parameters for (a) $O_{\mrm{APS}}$ (b) $O_{\mrm{BPS}}$, and (c) $O_{\mrm{HS}}$. The color boundaries correspond to the phase boundaries in Fig. 5.}
 \label{fig6}
\end{figure}

To distinguish the quantum phases, we introduce projection operators of singlet and triplet pairs of $i$-th and $j$-th sites given by,
\begin{equation}
  \mcl{P}_{i,j}^{s}=\frac{1}{4}-\vS_i\cdot\vS_j, \hspc \mcl{P}_{i,j}^{t}=\vS_i\cdot\vS_j+\frac{1}{4}.  
\end{equation}
We first define order parameters of the APS and BPS phases as products of these projection operators of paired spins on opposite apices of icosahedron, i.e., 
\begin{equation}
  O_{\mrm{APS}}=\ev{\prod_i \mcl{P}_{i,\bar{i}}^{s}}, \hspc O_{\mrm{BPS}}=\ev{\prod_i \mcl{P}_{i,\bar{i}}^{t}},
\end{equation}
where $\bar{i}$ denotes the opposite site from $i$-th apex of icosahedron. 
Figure 6 shows the order parameters in the parameter space. We can see that the APS and BPS phases are well distinguished, while $O_{\mrm{BPS}}$ gives a non-zero value, that is, the unity in the ferromagnetic phase. Note that the quantum state in the APS phase also includes triplet configurations, resulting in a non-zero value of the BPS order. On the other hand, these order parameters show zero in the HS and DHS phases. We thus consider a combination of singlet and triplet defined by
\begin{equation}
  O_{\mrm{HS}}=\ev{\prod_{(i,j)=(0,5),(1,6),(2,3)} \br{\mcl{P}_{i,\bar{i}}^{s}\mcl{P}_{j,\bar{j}}^{t}-\mcl{P}_{i,\bar{i}}^{t}\mcl{P}_{j,\bar{j}}^{s}}}.
\end{equation}
The site combinations $(i,j)=(0,5),(1,6)$, and $(2,3)$ correspond to the 1st neighbor bonds in three perpendicular planes of icosahedron (see Fig. 1). Therefore, the HS order parameter represents products of singlet-triplet configurations on three perpendicular planes. In Fig. 6(c), we can see two finite-value regions of the HS order, i.e., positive values for the HS phase and negative values for the DHS phase. The difference of sign corresponds to an exchange of singlet and triplet pairs. Site exchange of the singlet and triplet pairs on each plane induces an exchange of the first and second neighbor bonds between planes. Therefore, the HS order parameter can probe an asymmetry of the first and second neighbor bonds.

\section{Conclusion}
In this paper, we have investigated magnetic ground states in both classical and quantum Heisenberg spin models on an icosahedral cluster, where all bonds are considered as ferromagnetic or antiferromagnetic exchange interactions. 
The ground-state phase diagrams have been numerically determined by using simulated annealing and exact diagonalization methods. 
Moreover, we have shown analytical explanations of spin configurations at specific points in the parameter space. 
Based on the numerical and analytical examinations, we have characterized four ground-state phases, i.e., the HA, DHA, PPA, and F phases with the APA state in the classical model. 
On the other hand, we have also classified the ground-state phases in the quantum model with numerical results on the analogy of the classical phases. 
In fact, we have successfully demonstrated the qualitative coincidence between the classical and quantum phases. 
Furthermore, we have found a distinctive quantum phase, the APS phase, in addition to four quantum analogs of classical phases, namely the BPS, HS, DHS, and F phases together with those order parameters.
The icosahedral spin clusters are in general found in the Tsai-type quasicrystals and approximants. 
In these alloys, spins are coupled with each other via so-called the RKKY interactions, and therefore, the icosahedral spin clusters are not isolated but interact with each other. 
However, magnetic properties can strongly reflect characteristics of an isolated icosahedral spin cluster if intra cluster interactions are relatively large enough as compared with inter cluster interactions. 
Thus, our study can give a good starting point to understand the magnetic properties experimentally observed in the Tsai-type quasicrystals and approximants.

\begin{acknowledgments}
We would like to thank T. J. Sato and T. Hiroto for fruitful discussions. 
This work was partly supported by Challenging Research (Exploratory) (Grant No.JP17K18764), Grant-in-Aid for Scientific Research on Innovative Areas (Grant No.JP19H05821). 
\end{acknowledgments}

\end{document}